\documentclass[11pt]{article}
\usepackage{epsfig}
\usepackage{moriond,amssymb}

\def\be{\begin{equation}}
\def\ee{\end{equation}}
\def\bea{\begin{eqnarray}}
\def\eea{\end{eqnarray}}

\newcommand{\epem}{e^+e^-}
\newcommand{\hr}{\mathcal{H_R}}
\newcommand{\hc}{\mathcal{H_C}}
\newcommand{\as}{\alpha_\mathrm{s}}

\begin{document}
\newcommand{\titlefootnote}{}
%
\begin{flushright}
  LPTHE--P02--03\\
  hep-ph/0205161\\
  May 2002
  \vspace{-2.0cm}
\end{flushright}
\renewcommand{\titlefootnote}{\footnote{Talk presented at the XXXVIIth
    Rencontres de Moriond `QCD and high energy hadronic interactions',
    Les Arcs, France, and at the X International Workshop on Deep
    Inelastic Scattering (DIS2002) Cracow, Poland.}}
%
%
\vspace*{4cm}
\title{DIS EVENT-SHAPE RESUMMATIONS AND SPIN-OFFS\titlefootnote%
}

\author{ M. Dasgupta$^1$ and G.~P. Salam$^2\,$\footnote{Speaker}}

\address{$^1$DESY, Theory Group, Notkestrasse 85, Hamburg, Germany.\\
$^2$LPTHE, Universit\'es P. \& M. Curie (Paris VI) et Denis Diderot
  (Paris VII), Paris, France.}

\maketitle\abstracts{We present results from a recently completed
  project to calculate next-to-leading logarithmic resummed
  distributions for a variety of event shapes in the 1+1-jet limit of
  DIS. This allows fits for the strong coupling and for
  non-perturbative effects using the large amount of data on these
  observables from HERA.  Spin-offs include the discovery of a new
  class of logs for certain final state observables (non-global
  observables); a program that allows a speed-up by an order of
  magnitude of certain fixed-order calculations in DIS with DISENT or
  DISASTER++; and the development of state-of-the-art PDF evolution
  code.}

\section{Introduction}

Event-shapes are observables sensitive to the flow of energy and
momentum in hadronic final states. They have been extensively studied
in $\epem$ collisions, for example for the measurement of the strong
coupling, tests of QCD through fits for the colour factors and the
study of novel approaches to hadronization.\cite{BethkeReview}
Typically the most discriminatory studies make use of event-shape
\emph{distributions}, which are compared to next-to-leading
perturbative predictions that are resummed in the $2$-jet limit.

Recently the HERA experiments have also started considering event
shapes, defined in the current hemisphere of the Breit frame. In
particular, distributions have been measured by H1,\cite{H1dist} and
while only mean values have been so far studied by
ZEUS~\cite{ZEUSmeans} it is our understanding that they intend to
extend their studies to distributions.

Resummed predictions for $\epem$ event shapes have existed in the
literature since the early nineties,\cite{eeResum} but until recently
no such calculations were available for DIS observables (an exception
is jet rates~\cite{DISjetrates}). Because of the strong similarity
between a hemisphere of an $\epem$ event and the current hemisphere of
the DIS Breit frame, it is natural to assume that the extension from
$\epem$ to DIS will be fairly straightforward. It turns out not to be
so, for both conceptual and technical reasons. In what follows we
outline some of the issues that arise, and then present some
preliminary comparisons to data.

\section{Resummation issues in DIS}

The most obvious new issue to arise in DIS compared to $\epem$ is that
of collinear factorization. Despite the fact that the observables are
all defined in the current-hemisphere ($\hc$) of the Breit frame,
owing to details of the kinematics those defined with respect to the
photon axis are sensitive to emissions in the remnant hemisphere
($\hr$) through recoil effects. Requiring the event-shape to have a
value close to that of the $1+1$-jet limit, one therefore forbids
emissions in the whole of the phase space ($\hc$ and $\hr$). However
collinear factorization at the scale $Q^2$ is conditional on there
being no restrictions on emissions in the remnant hemisphere.
Requiring the event shape to have a value less than some $V$, which
translates to a limit on the largest possible transverse momentum of
collinear emissions in $\hr$, $k_t^2 \lesssim V^n Q^2$ ($n$ is
observable-dependent), has the consequence~\cite{ADS} that collinear
factorization can only be recovered if parton distributions are
evaluated at a scale of the order of $V^n Q^2$. This is actually a
familiar result from calculations of the $p_t$ distribution of
Drell-Yan pairs~\cite{DYpairs} and has also been observed in more
complicated multi-jet event shapes.\cite{BMSZ}

\begin{figure}[hbt]
    \epsfig{file=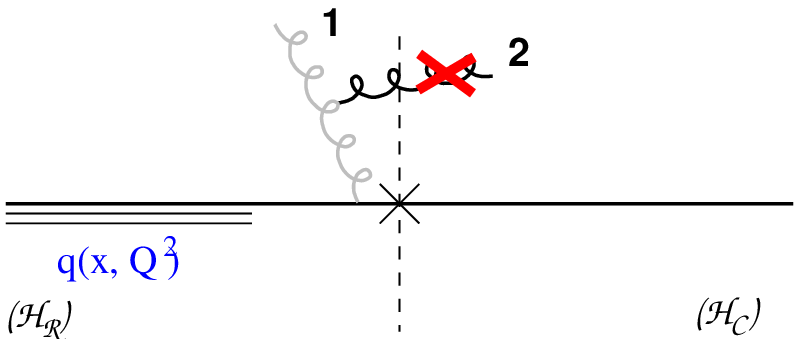,width=0.48\textwidth}\hfill
    \epsfig{file=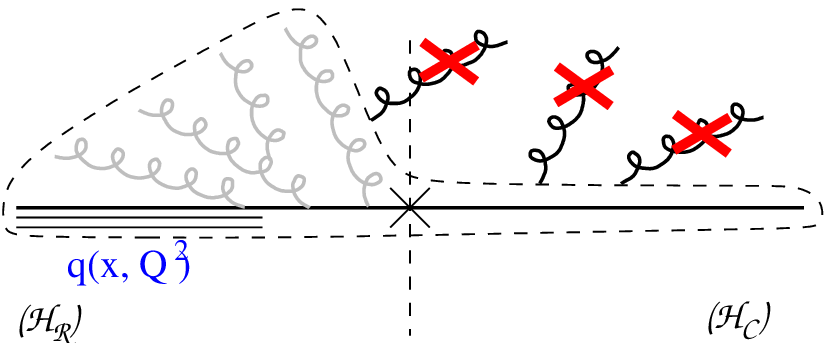,width=0.48\textwidth}
    \caption{Contributions relevant in the calculation of non-global
      terms; the triple lines indicate incoming partons.}
    \label{fig:nonglobal}
\end{figure}

A second issue is that of non-global logarithms, which arise in
observables sensitive only to emissions in a restricted portion of
phase space (e.g.\ $\hc$) such as the jet mass, and additionally in
observables whose sensitivity to emissions is discontinuous across one
or more boundaries in phase space (an example~\cite{DSinprep} is the
thrust $\tau_{zE}$). An erroneous assumption that has widely been made
in the literature~\cite{ADS,BDMZ,BG,SUNY} is that (to
single-logarithmic accuracy) in order to suppress radiation into $\hc$
(say), it suffices to suppress \emph{primary} radiation from the
various hard `legs' into that hemisphere.

While at leading order in $\as \ln V$ this is correct, starting from
second order, configurations such as those in fig.~\ref{fig:nonglobal}
become relevant.~\cite{DSng} The crosses indicate emissions which must
be forbidden in order for the observable to have a small value. The
grey emissions are those that do not directly affect the value of the
observable. The left-hand picture represents the configuration
relevant at second order: a soft emission (1) in $\hr$, which does not
contribute to the observable, radiates an even softer emission (2) into
$\hc$, which does contribute to the observable. The strong ordering in
energies $Q\gg E_1 \gg E_2$ leading to one power of $\ln V$ for each
power of $\as$.  While this term is calculable analytically, at all
orders one needs to forbid coherent radiation into $\hc$ from
arbitrarily complicated ensembles of large-angle energy-ordered gluons
in $\hr$ (right-hand picture).  This is complicated both from the
point of view of the colour structure and of the geometry.  The former
can be dealt with approximately in the large $N_C$ approximation,
while the latter can so far only be treated numerically.  Some insight
into the dynamics associated with these non-global logs was obtained
in the context of a more general study of energy flow
distributions,\cite{DSEflow} where one finds that in the limit of
large $\ln 1/V$ not only is radiation into $\hc$ forbidden, but
radiation at intermediate energy scales is also forbidden in a
neighbouring `buffer' region of $\hr$. The size (in rapidity) of this
buffer region increases with $\ln V$ and the overall suppression
factor coming from non-global logs seems, at least in part, to be
associated with the suppression of primary radiation into the buffer.

\section{Technical issues}

When implementing the resummations as computer programs to allow
comparisons to data a number of technical issues arose. When this
project started there existed only two subtraction-based programs for
NLO calculations in DIS, namely DISENT~\cite{DISENT} and
DISASTER++~\cite{DISASTER}, which were known to disagree for certain
observables.~\cite{DISdisagree} Comparisons with the expansion of the
resummed results made it possible to identify DISASTER++ as the one
giving correct predictions.\footnote{The recently released NLOJET
  program~\cite{NLOJET} is also in agreement with DISASTER++.}

Unfortunately, of the two programs, DISASTER++ is an order of
magnitude slower, and we would have needed over a year's computing
time to obtain the fixed-order predictions (needed to describe the
observable in the region outside the $1+1$ jet limit) to sufficient
accuracy. However, the traditional approach to such calculations
involves considerable duplication of effort: one needs distributions
at several $x$ and $Q^2$ values and typically one uses separate events
for each $x$ and $Q^2$ value. But the matrix elements (modulo their
$y_{Bj}$-dependence) and the calculation of the event-shape are both
independent of $x$ and $Q^2$, so  each event in an NLO Monte-Carlo
program can be `reapplied' to several $x$ and $Q^2$ values. We have
written a program DISPATCH,~\footnote{Available from
  \texttt{http://www.lpthe.jussieu.fr/\~~$\!\!\!$salam/dispatch/}}
which acts as a wrapper to DISENT and DISASTER++ so as to automate
such a procedure.

Another spin-off from this project is the development of a
high-precision PDF evolution code,\cite{DSBroad} which has been used
in collaboration with Vogt~\cite{SV} to produce reference NNLL
evolutions to an accuracy of 1 part in $10^5$.

\section{Comparison to data}

\begin{figure}[t]
  \epsfig{file=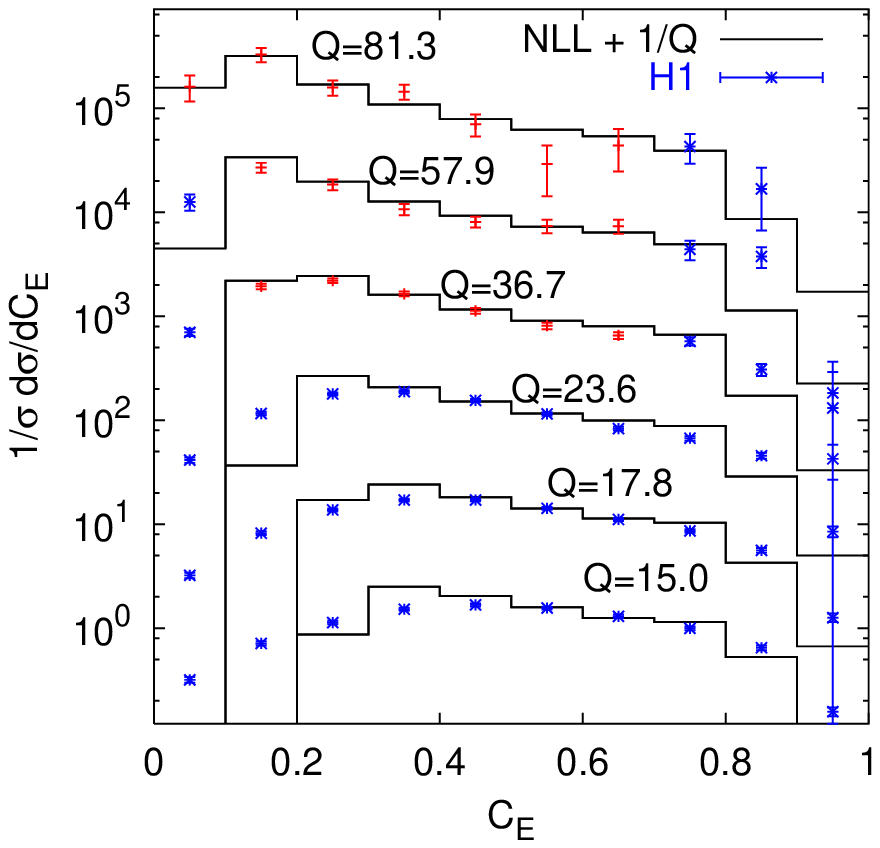,width=0.48\textwidth}\hfill
  \epsfig{file=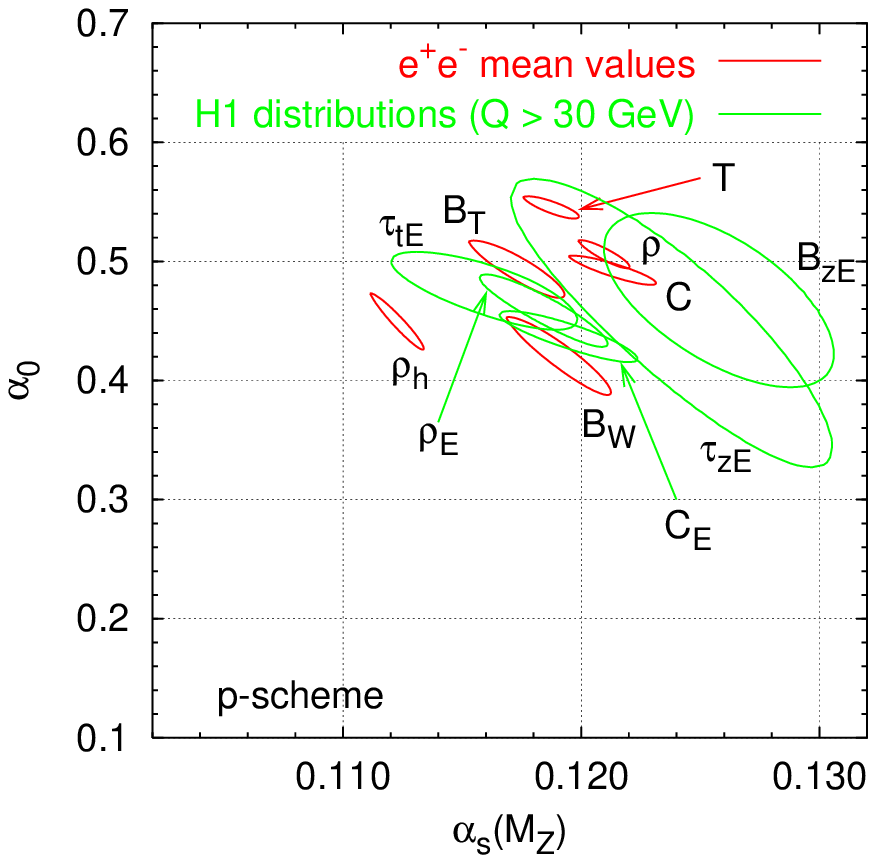,width=0.48\textwidth}
  \caption{left, a comparison between resummed predictions and H1
    results; right, results of fits for $\as$ and $\alpha_0$.}
  \label{fig:results}
\end{figure}

The left-hand plot of fig.~\ref{fig:results} shows a comparison
between our matched resummed distributions~\cite{ADS,DSinprep,DSBroad}
and the H1 data~\cite{H1dist} for the $C$-parameter, $C_E$.
Non-perturbative contributions have been included using $1/Q$
corrections, whose size have been hypothesized~\cite{OneOverQ} to be
governed by a universal parameter $\alpha_0$. We have fitted for both
$\as$ and $\alpha_0$, using only the red points of the distributions.
The results ($1$-$\sigma$ contours) for $C_E$ and a number of other
DIS observables are shown in the right-hand plot (green curves) and
compared to $\epem$ results for mean values (red curves), with the jet
masses measured in so-called massless schemes.~\cite{SW} The agreement
both within DIS and across experiments is strong confirmation of the
universality hypothesis for $\alpha_0$.

There remain some observables in DIS where the agreement is less good,
notably those measured with respect to the photon axis ($\tau_{zE}$,
$B_{zE}$), and the situation worsens if one includes lower $Q$ (also
lower $x$) data. The detailed origin of the problem remains to be
understood, though it may well be associated with higher-order
corrections being relatively larger at lower $Q$ values. In this
context, higher precision data (and higher resolution data as well),
especially at larger $Q$ values, will be interesting as it will make
it possible to pin down any systematic $Q$ dependence over and above
that expected from the theoretical predictions used so far, perhaps
for example from shape functions.\cite{ShapeFunctions}

\end{document}